\documentclass[twocolumn]{svjour3}          
\smartqed  
\usepackage{graphicx}
\usepackage{algorithm}
\usepackage{algpseudocode}
\usepackage{cite}                      %
\usepackage{epsfig}
\usepackage{graphics,color}
\usepackage{verbatim}
\usepackage{times,amsmath}
\usepackage{multirow}
\usepackage{cases}
\usepackage[tight]{subfigure}
\usepackage{verbatim}
\usepackage{amssymb}
\usepackage{comment}
\usepackage[titletoc]{appendix}

\DeclareMathOperator*{\E}{E}
\begin{document}

\title{Estimating Distances via Received Signal Strength and Connectivity in Wireless Sensor Networks}


\author{Qing Miao, Baoqi Huang$^*$ and Bing Jia}


\institute{Inner Mongolia University, Hohhot, 010021, China,\\
              \email{13734812507@163.com, cshbq@imu.edu.cn and jiabing@imu.edu.cn}        }

\date{Received: date / Accepted: date}

\maketitle

\begin{abstract}
Distance estimation is vital for localization and many other applications in wireless sensor networks (WSNs). Particularly, it is desirable to implement distance estimation as well as localization without using specific hardware in low-cost WSNs. As such, both the received signal strength (RSS) based approach and the connectivity based approach have gained much attention. The RSS based approach is suitable for estimating short distances, whereas the connectivity based approach obtains relatively good performance for estimating long distances. Considering the complementary features of these two approaches, we propose a fusion method based on the maximum-likelihood estimator (MLE) to estimate the distance between any pair of neighboring nodes in a WSN through efficiently fusing the information from the RSS and local connectivity. Additionally, the method is reported under the practical log-normal shadowing model, and the associated Cramer-Rao lower bound (CRLB) is also derived for performance analysis. Both simulations and experiments based on practical measurements are carried out, and demonstrate that the proposed method outperforms any single approach and approaches to the CRLB as well.
\end{abstract}


\section{INTRODUCTION}
Wireless sensor networks (WSNs), composed of hundreds or thousands of small and inexpensive nodes with constrained computing power, limited memories, and short battery lifetime, can be used to monitor and collect data in a region of interests. Accurate and low-cost node localization is important for various applications in WSNs, and thus, great efforts have been devoted to developing various localization algorithms, categorized into distance based algorithms and connectivity-based algorithms \cite{mao2007wireless}. The distance-based localization algorithms rely on distance estimates and are able to achieve relatively good localization accuracy, whereas the connectivity-based localization algorithms generally achieve coarse-grained localization accuracy since only local connectivity information (the numbers of the common and non-common one-hop neighbors) is employed for distance estimation. Besides, distance estimation is also useful for sensor network management, such as topology control \cite{li2005cone,siripongwutikorn2008mobility} and boundary detection \cite{sitanayah2010heuristic,Huang2013LCN}.

In reality, distance estimation can be realized by using information such as RSS, time of arrival (TOA) and time difference of arrival (TDOA) \cite{mao2007wireless}. The RSS approach (using RSS measurements) does not require any dedicated hardware, but is able to provide coarse-grained distance estimates; in contrast, the TOA and TDOA methods can provide distance estimates with high accuracy at the cost of extra hardware, but it is unaffordable to equip each sensor with a dedicated measurement device in a large-scale WSN due to the costs in both hardware and energy.  Therefore, it is of great importance to enhance the accuracy of low-cost distance measurement approaches, and many efforts have been imposed in the literature.  Due to its intrinsic simplicity and independence of dedicated hardware, the RSS-based distance methods have gained much attraction \cite{Chitte2009,Benkic2008Using,Heurtefeux2012,Rosa2012Human}. But, both simulations and theoretical analysis indicate that the performance of the RSS-based methods is relatively poor, and degrades with the increasing actual distances \cite{Benkic2008Using,Chitte2009,huang2014estimating}. Hence, it is necessary to advance the RSS-based methods by adopting various techniques \cite{Botta2013Adaptive,Zanella2014RSS}.

Apart from RSS measurements, local connectivity information of each node is also independent of extra device, and can be employed to estimate distances from itself to other neighboring nodes \cite{nagpal2003organizing,shang2003localization,huang2014estimating}. However, Unlike the RSS-based methods that return ideal estimates for short distances, the connectivity-based methods obtain relatively good performance for estimating long distances. Therefore, the complementary features of these two types of distance estimation methods motivate us to design a fusion method which is able to sufficiently exploit the advantages of both types of methods.



In this paper, the maximum-likelihood estimator (MLE) is utilized to efficiently fuse the information from the RSS measurement and local connectivity, so as to provide good performance regardless the actual sizes of distances to be estimated. The strengths of the proposed fusion method lie in the following aspects. Firstly, it is well known that the MLE is asymptotically optimal, indicating that the proposed method is able to obtain superior performance. Secondly, the practical log-normal (shadowing) model is adopted to characterize the probabilistic distributions of the errors respectively induced by the RSS-based method and the connectivity-based method, so as to ensure the applicability of the proposed method. Thirdly, the Cramer-Rao lower bound (CRLB) associated with the proposed method is also formulated and can be used to evaluate the optimality of the proposed method. Finally, both experiments based on measurements in a real environment and simulations are carried out to thoroughly validate the effectiveness of the proposed method. However, even if the proposed method outperforms the other two available methods and approaches to the CRLB in most time, it still suffers from a few limitations: it is computationally expensive due to the complicated cost function involved; as the connectivity-based method, the performance also relies on the sensor density. This study not only contributes to improving sensor localization in low-cost WSNs, but also paves the way for advancing many other researches and applications relying on inter-node distances.

The remainder of the paper is organized as follows. Section \ref{related} briefly reviews related works in the literature. Section \ref{sec:model} introduces the WSN model and both the  RSS-based and connectivity-based distance estimation methods. Section \ref{sec:fusion} presents the models of both the RSS-based and connectivity-based methods, and proposes the fusion method based on the MLE and formulates the corresponding CRLB. Section \ref{sec:exp} reports the performance of the proposed method through both simulations and experiments. Finally, section \ref{sec:con} concludes the paper and sheds lights on future works.

\section{RELATED WORKS}\label{related}
In this section, we shall briefly review the studies on distance estimation in low-cost WSNs, which can be categorized into RSS-based methods and connectivity-based methods.

The RSS-based methods infer distances from power losses incurred by signals travelling between transmitter and receiver as long as after the model depicting the relationship between power losses and distances is available. In \cite{Chitte2009}, an estimator was designed based on the MLE and the log-normal model, but the theoretical analysis indicated that the estimator is inefficient in the sense that the error variance increases exponentially with powers. However, the practice in \cite{Benkic2008Using,Heurtefeux2012} reveal that the RSS-based distance estimation is unreliable. Additionally, the distance estimation is even complicated in indoor environments since the factors, like furniture, hand grip and human bodies, affect the distance estimation \cite{Rosa2012Human}. Hence, in \cite{Botta2013Adaptive} a dynamic calibration method was proposed to update the log-normal model parameters which fluctuate with environmental changes, and in \cite{Zanella2014RSS}, an averaging method based on multichannel RSS measurements was presented to mitigate the variability of RSS measurements. Therefore, it can be concluded that it is still challenging to apply the RSS-based methods.

The connectivity-based methods infer distances from local connectivity information among different nodes in WSNs \cite{buschmann2006estimating,buschmann2007radio,huang2014estimating}. The neighborhood intersection distance estimation scheme (NIDES) presented in \cite{buschmann2006estimating} heuristically relates the distance, e.g. from node A to node B, to an easily observed ratio, i.e. the number of their common immediate neighbors to the number of immediate neighbors of A, and then performs distance estimation at node A according to this ratio and other a priori known information. NIDES assumes a unit disk model, namely that the communication coverage of each node is a perfect disk, and all nodes are uniformly and randomly deployed in the WSN. Its enhanced version presented in \cite{buschmann2007radio} adapted the ratio by taking into account the number of immediate neighbors of node B, and heuristically stated that NIDES could be applied under arbitrary communication models. In \cite{huang2014estimating}, a novel method is presented based on the MLE under a generic channel model, including the unit disk model and the more realistic log-normal model, and its error characteristics were analyzed in light of the CRLB. However, the performance of the connectivity-based method obtains obvious errors when estimating short distances.

In summary, both of the above methods are restricted in practical applications, but fortunately, are complementary to each other, which motivates us to combine them to obtain better performance. As such, this paper presents a fusion method to estimate distances by making use of RSS measurements and local connectivity under the practical log-normal model.

\section{PRELIMINARIES}\label{sec:model}
This section first briefly introduces the static WSN model which is considered in this paper, and then elaborates the RSS-based and connectivity-based methods, respectively. Throughout this paper, we shall use the following mathematical notation: $p(\cdot)$ denotes the probability density function of an event, and $\E(\cdot)$ denotes the statistical expectation.

\subsection{The WSN Model}

In a static WSN, nodes are often assumed to be randomly and uniformly distributed on account of the random nature of network deployment, e.g. nodes being dropped from a flying plane. Since a homogeneous Poisson process provides an accurate model for the uniform distribution of nodes as the network size approaches infinity, we define the static WSN to be deployed over an infinite plane according to the homogeneous Poisson process of intensity $\lambda$.

\subsection{The RSS-based Method}\label{sec:RSSmodel}

The RSS-based method estimates the distance between any pair of nodes using the received signal power, i.e. RSS. When a signal is propagated between transmitter and receiver, the power loss or attenuation is unavoidable, and generally rises with increasing the separation between  transmitter and receiver. Moreover, as is commonly made in both theoretical studies (e.g. \cite{li2007collaborative,ouyang2010received,so2011linear}) and experimental studies (e.g. \cite{patwari2003relative,cheng2009new}), the power loss can be formulated by using the log-normal model, namely
\begin{equation}\label{eq:log-normal}
    P_R(d)\text{(dBm)}=\overline{P_R}(d_0)\text{(dBm)}-10\alpha \log_{10}\frac{d}{d_0}+Z,
\end{equation}
where $P_R(d)$(dBm) is the received signal power at $d$ in dBm, $\overline{P_R}(d_0)$(dBm) is the mean received signal power at a reference distance $d_0$ in dBm, $\alpha$ is the path loss exponent, and $Z$ is a random variable representing the shadowing effect, normally distributed with mean zero and variance $\sigma_{dB}^2$.

Based on the log-normal model in (\ref{eq:log-normal}), it is straightforward to infer the distance $d$ from the received signal power $P_R(d)$ by using any parameter estimator. For instance,  $\hat{d_R}$ is defined to be the distance estimate between two nodes via an associated RSS measurement, and can be formulated as follows (see \cite{BottaM2013} for more details)
\begin{equation}\label{eq:RSSdis}
\hat{d_R}=10^{\frac{\overline{P_R}(d_0)(dBm)-P_R(d)(dBm)}{10\alpha}}d_0.
\end{equation}

\subsection{The Connectivity-based Method}\label{sec:distancemodel}
The connectivity-based method estimates the distance between any pair of neighboring nodes on the basis of their local connectivity information. In this subsection, we present this method under both the simple unit disk model and the generic channel model. Specifically, the unit disk model assumes an ideal communication coverage for each node, i.e. a perfect disk with the radius of $r$, whereas the generic channel model, including the log-normal model, takes into consideration the random noises (e.g. the shadowing effect) in the communication channels, so as to characterize the communication coverage in a more practical way.

\subsubsection{The Unit Disk Model Case}\label{subsec:unit}
\begin{figure}
  \centering
  \includegraphics[width=0.5\textwidth]{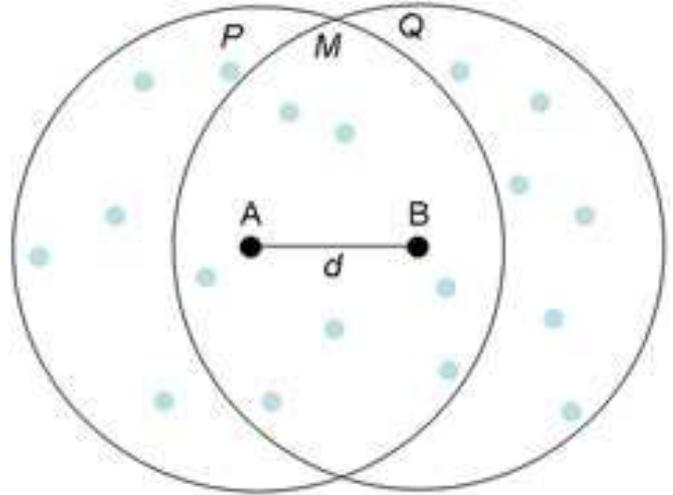}
  \caption{The communication coverage of two nodes under the unit disk model.}
  \label{fig:network}
\end{figure}

Given a static WSN, suppose two nodes A and B with coordinates $(x_A,y_A)$, $(x_B,y_B)$ and separation $d$ ($d \leq r$) and the disks with the common radius $r$ represent their individual communication coverage under the unit disk model, as shown in Fig. \ref{fig:network}. Because of $d\leq r$, the two disks intersect and create three disjoint regions. Regarding $r$ as  constant, define $S=\pi r^2$ and $f(d)$ to be the area of the middle region in Figure \ref{fig:network}, where
\begin{equation}
f(d)=\frac{2S}{\pi}\arccos\left(\frac{d}{2r}\right)-d\sqrt{r^2-\frac{d^2}{4}}.
\end{equation}

It is obvious that the nodes residing in the middle region are common immediate neighbors of A and B, the nodes residing in the left (or right) one are non-common immediate neighbors of A (or B). Define three random variables $M$, $P$, and $Q$ to be the numbers of the three categories of neighbors; according to the assumption of the Poisson point process, they are mutually independent and Poisson with means $\lambda f(d), \lambda (S-f(d))$ and $\lambda (S-f(d))$, as pointed out in \cite{franceschetti2008random}. However, the actual values of $M$, $P$, and $Q$ can be easily obtained after A and B exchange their neighborhood information. On the basis of the observations of $M$, $P$ and $Q$ and the method of MLE, the distance estimate of $d$, denoted $\hat{d_c}$, can be summarized as follows (see \cite{huang2014estimating} for details)
\begin{equation}
\hat{d_c}=\left \{
              \begin{aligned}
                &f^{-1}(S),            & &\mathbf{if} \quad M=P=Q=0;  \\
                &f^{-1}(\hat{\rho}S),  & &\mathbf{otherwise}   \\
              \end{aligned}
        \right.
\end{equation}
where $\hat{\rho}=\frac{2M}{2M+P+Q}$.

\subsubsection{The Generic Channel Model Case}

In the generic channel model \cite{huang2014estimating}, the randomness on the RSS can be characterized by a function $g(d)$, denoting the probability that a directional communication link exists from transmitter to receiver with distance $d$. In particular, in the log-normal model, we can have
\begin{equation}\label{eq:g(d)}
g(d)=\int^{\infty}_{k\log\frac{d}{r}}\frac{\exp-\frac{z^2}{2\sigma_{dB}^2}}{\sqrt{2 \pi}\sigma_{dB}}dz
\end{equation}
where $k=10\alpha/\log10$; $r$ denotes a pseudo transmission range which depends on the antenna gains, the wavelength of the propagating signal, the transmission power and the communication threshold for RSS.

Let $M$, $P$, and $Q$ continuously denote the numbers of common and non-common immediate neighbors associated with two nodes. we can compute their expectations as follows
\begin{equation}
 \begin{aligned}
 &E(M+P)=E(M+Q)=\\
        &\lambda \int_{-\infty}^\infty \int_{-\infty}^\infty  \times g(\sqrt{(x-x_B)^2+(y-y_B)^2})dxdy,
\end{aligned}
\end{equation}

 \begin{equation}
 \begin{aligned}
 E(M)=&\lambda \int_{-\infty}^\infty \int_{-\infty}^\infty g(\sqrt{(x-x_A)^2+(y-y_A)^2}) \\
      & \times g(\sqrt{(x-x_B)^2+(y-y_B)^2})dxdy.
\end{aligned}
\end{equation}

Then, by generalizing $S$ and $f(d)$ to specify the expectations of $M$, $P$ and $Q$ under the generic channel model instead of the areas defined under the unit disk model, we can have the following formulas
\begin{equation}
\begin{aligned}
S=\int_{-\infty}^\infty \int_{-\infty}^\infty g(\sqrt{(x-x_B)^2+(y-y_B)^2})dxdy,
\end{aligned}
\end{equation}

\begin{equation}\label{eq:f(d)}
 \begin{aligned}
 f(d)=&\int_{-\infty}^\infty \int_{-\infty}^\infty g(\sqrt{(x-x_A)^2+(y-y_A)^2}) \\
   & \times g(\sqrt{(x-x_B)^2+(y-y_B)^2})dxdy.
\end{aligned}
\end{equation}

Moreover, by using (\ref{eq:g(d)}) and (\ref{eq:f(d)}), we can derive the formula for $f(d)$ under the log-normal model. Similar to \ref{subsec:unit}, the distance estimate can be calculated based on the inverse of $f(d)$; that is,
\begin{equation}\label{eq:conectDis}
\hat{d_c}=\left \{
              \begin{aligned}
                &0,            & &\mathbf{if} \quad  M=P=Q=0;  \\
                &f^{-1}(\hat{\rho}S),  & &\mathbf{if}  \quad f(d_{th})\leq \hat{\rho}\leq f(0);     \\
                & d_{th},      & & \mathbf{if} \quad \hat{\rho}S< f(d_{th})\\
              \end{aligned}
        \right.
\end{equation}
where $\hat{\rho}=2M/(2M+P+Q)$, and $d_{th}$ denotes the longest distance between two neighboring nodes.

However, since the closed-form formulae for $f(d)$ and its inverse are hard or even impossible to obtain, we thus substitute its inverse by using an approximate piecewise linear function, namely that a linear regression model is established to predict $d$ for each affine segment.

\section{The Proposed Fusion Method}\label{sec:fusion}
In this section, we introduce the new distance estimation method based on the aforementioned RSS-based method and connectivity-based method. To do so, it is necessary to understand the statistical distributions of the RSS-based and connectivity-based distance estimates, respectively. As such, a thorough theoretical analysis is carried out to investigate both of the distance estimation methods. After that, the MLE method can be instantly applied to estimate the distance, and particularly, the Newton-Raphson method is adopted to solve the corresponding likelihood function. Besides, the CRLB associated with the proposed fusion distance estimation method is formulated to further observe its performance.

\subsection{Statistical Analysis of Distance Estimates}
In what follows, the RSS-based and connectivity-based distance estimation methods shall be analyzed by formulating their statistical distributions, which paves the way for clarifying the proposed distance estimation method.

\subsubsection{The RSS-based Case}

In light of the RSS-based distance estimation method presented in Subsection \ref{sec:RSSmodel}, it follows from Equation (\ref{eq:log-normal}) that $d$ can be formulated as follows
\begin{equation}
d=10^{\frac{\overline{P_R}(d_0)(dBm)-P_R(d)(dBm)}{10\alpha}} \times 10^{\frac{Z}{10\alpha}}.
\end{equation}
By replacing the RSS-based distance estimate $\hat{d_R}$ in (\ref{eq:RSSdis}), we can have
\begin{equation}\label{eq:RSSnoiseDis}
d=\hat{d_R}\times 10^{\frac{Z}{10\alpha}}.
\end{equation}
Then, define $\epsilon_R$ to be the multiplicative error of the RSS-based distance estimate, namely
\begin{equation}\label{eq:RSShatdr}
\hat{d_R}=d\times \epsilon_R,
\end{equation}
and
\begin{equation}\label{eq:RSSnoise}
\epsilon_R=10^{-\frac{Z}{10\alpha}}.
\end{equation}

Since $\epsilon_R$ is dependent on the normal variable $Z$, their probability density functions, denoted by $p_{\epsilon_R}(\cdot)$ and $p_Z(\cdot)$ respectively, satisfy the following equation
\begin{equation}\label{eq:functionth}
p_{\epsilon_R}(x)=p_Z(z_{\epsilon_R}(x))\times |z_{\epsilon_R}'(x)|,
\end{equation}
where
\begin{equation}\label{eq:RSSnoise}
z_{\epsilon_R}(x)=-10\alpha\times\log_{10}x.
\end{equation}
Then, we can obtain the probability density function of $\epsilon_R$ as follows
\begin{equation}\label{eq:RSSnoiseD}
p_{\epsilon_R}(x)=\frac{1}{\sqrt{2\pi}\sigma_R}\exp\left(-\frac{\log_{10}^2{x}}{2\sigma_R^2}\right)\frac{1}{x\ln10},
\end{equation}
where $\sigma^2_R=\frac{\sigma_{dB}^2}{(10\alpha)^2}$.

Since the relationship of $\epsilon_R$ and $\hat{d_R}$ is expressed as (\ref{eq:RSShatdr}) and the above method, the probability density function of the RSS-based distance estimation, denoted by $p_{\hat{d_R}}(\cdot)$, satisfies the following equation
\begin{equation}\label{eq:Rdispro}
 p_{\hat{d_R}}(x)=\frac{1}{\sqrt{2 \pi}\sigma_R}\exp\left(-\frac{\log_{10}^2\frac{x}{d}}{2\sigma_R^2}\right)\frac{1}{x\ln10}.
\end{equation}


\subsubsection{The Connectivity-based Case}


In Subsection \ref{sec:distancemodel}, we have introduced the connectivity-based distance estimation method and a piecewise linear function to approximate the function $f(d)$. Similarly, we can approximate $f(d)$ by
\begin{equation}
f(d)\approx kd+b,
\end{equation}
where $k$ and $b$ are constant. Thus according to (\ref{eq:conectDis}), the connectivity-based distance estimate $\hat{d_c}$ satisfies
\begin{equation}
\hat{\rho}S=k\hat{d_c}+b,
\end{equation}
where $\hat{\rho}=2M/(2M+P+Q)$ is a random variable.

Then, define $\epsilon_c $ to be the error of the connectivity-based distance estimate, namely
\begin{equation}\label{eq:Cerdisrelation}
 \epsilon_c=\hat{d_c}-d,
\end{equation}
and
\begin{equation}\label{eq:noise2}
\epsilon_c=\frac{1}{k}(\hat{\rho}S-f(d)).
\end{equation}

Then, we are interested in the distribution of the error base on the formula of $\epsilon_c$. Because of the variable $M$, $P$ and $Q$ are mutually independent Poisson random variables with means $\lambda f(d)$, $\lambda (S-f(d))$, $\lambda (S-f(d))$, respectively, and the additivity property of the independent Poisson random variables. The distribution of $2M+P+Q$ can be formulated as
\begin{equation}
2M+P+Q \sim P(2 \lambda S).
\end{equation}

In \cite{griffin1992distribution}, it has proofed that the Poisson random variable with the mean $\lambda$ larger than five can approximately equal to a normal distribution with the mean and variance are equal to $\lambda$. The Poisson random variables $2M$ and $2M+P+Q$ with means $2\lambda f(d)$ and $2\lambda S$ satisfy above condition, thence the Poisson random variables can approximately be presented as follows

\begin{equation}\label{eq5}
2M \sim N(2\lambda f(d),2\lambda f(d)),
\end{equation}
\begin{equation}\label{eq6}
2M+P+Q \sim N(2\lambda S, 2\lambda S).
\end{equation}


Next, in order to analyse the distribution of the $\hat{\rho}=2M/(2M+P+Q)$, we consider the ratio of two independent normal random variables. In \cite{diaz2013existence}, it has proofed that the two independent normal variables $X$ and $Y$ with means and variances ($\mu_x$, $\sigma^2_x$) and ($\mu_y$, $\sigma^2_y$), respectively. The random variable $Z=X/Y$ could be approximated to the normal distribution with the mean and variance
\begin{equation}\label{eq:normal}
\begin{split}
\mu_z&=\frac{\mu_x}{\mu_y},\\
\sigma^2_z=\left(\frac{\mu_x}{\mu_y}\right)^2&\left(\left(\frac{\sigma_x}{\mu_x}\right)^2+\left(\frac{\sigma_y}{\mu_y}\right)^2\right).
\end{split}
\end{equation}
By (\ref{eq5}), (\ref{eq6}) and (\ref{eq:normal}), we can have the statistical distribution of $\hat{\rho}$ satisfies that
\begin{equation} \label{eq3}
   \hat{\rho} \sim N\left(\frac{f(d)}{S},\frac{f^2(d)}{S^2}\left(\frac{1}{2\lambda f(d)}+\frac{1}{2\lambda S}\right)\right).
\end{equation}

According to the formal in (\ref{eq:noise2}), we can obtain the error distribution of the connectivity-based distance estimation when $S$ and $f(d)$ are constant
\begin{equation}
   \epsilon_c \sim N(0,\sigma_c^2),
\end{equation}
where $\sigma_c^2=\frac{f^2(d)}{k^2}\left(\frac{1}{2\lambda f(d)}+\frac{1}{2\lambda S}\right)$.

Therefore, the probability density function of the error $\epsilon_c$, denoted by $p_{\epsilon_c}(\cdot)$, satisfies the following equation
\begin{equation}\label{eq:noiseDis2}
p_{\epsilon_c}(x)=\frac{1}{\sqrt{2 \pi}\sigma_c}\exp\left(-\frac{x^2}{2\sigma_c^2}\right),
\end{equation}
and then, the probability density function, denoted by $p_{\hat{d_c}}(\cdot)$, satisfies the following equation
\begin{equation}\label{eq:Cdispro}
 p_{\hat{d_c}}(x)=\frac{1}{\sqrt{2 \pi}\sigma_c}\exp\left(-\frac{(x-d)^2}{2\sigma_c^2}\right).
\end{equation}

\subsection{The Proposed Distance Estimation Method}
Given the error distributions of the RSS-based and connectivity-based distance estimates, the MLE can be applied to fuse the distance estimates by the above two methods. Since the distance estimates $\hat{d_R}$ and $\hat{d_c}$ rely on different sources of information, we assume that $\hat{d_R}$ and $\hat{d_c}$ are independent from each other and express the likelihood function as follows
\begin{equation}\label{eq:l}
\begin{split}
L(d)&= p_{\hat{d_R}}(x_1)\times p_{\hat{d_c}}(x_2)\\
 &=\frac{1}{2\pi\sigma_R\sigma_c}\exp\left(-\frac{\log_{10}^2\frac{x_1}{d}}{2\sigma_R^2}-\frac{(x_2-d)^2}{2\sigma_c^2}\right)\frac{1}{x_1\ln10}.\
\end{split}
\end{equation}
Then, the natural logarithm of the likelihood function is
\begin{equation}
\ln L= -\ln (2\pi\sigma_R\sigma_cx_1\ln10)-\frac{\log_{10}^2\frac{x_1}{d}}{2\sigma_R^2}-\frac{(x_2-d)^2}{2\sigma_c^2}.
\end{equation}

In order to obtain the maximum value of $\ln L$, the Newton-Raphson method is adopted to derive the root of the first derivative of $\ln L$, denoted by
\begin{equation}
F(\hat{d})=\frac{\log_{10}\frac{x_1}{\hat{d}}}{\sigma_R^2\ln10}+\frac{\hat{d}(x_2-\hat{d})}{\sigma_c^2}.
\end{equation}
To do so, let $F'(\hat{d})$  be the derivative function of $F(\hat{d})$, and the specific steps of successively finding better approximations to the root of the function $F(\hat{d})$ include
\begin{enumerate}
  \item Select an initial guess $d_0$, where $\hat{d_0}=(x_1+x_2)/2$;
  \item Calculate the values of the function $F(d_k)$ and derivative $F'(d_k)$;
  \item Update $\hat{d}_{k+1}$ by using the iterative equation $\hat{d}_{k+1}=\hat{d}_k-\frac{F(\hat{d}_k)}{F'(\hat{d}_k)}$;
  \item Repeat Step 2 and 3 until $|\hat{d}_{k+1}-\hat{d}_k|<\xi$, where $\xi$ is a sufficiently small positive number.
\end{enumerate}

\subsection{CRLB}
The CRLB expresses a lower bound on the variance of any unbiased estimator \cite{salman2011effects,ling2012quantification} and is equal to the inverse of the corresponding Fisher Information Matrix (FIM). In this subsection, the CRLB regarding the proposed distance estimation problem, namely estimating the distance $d$ from $M$, $P$, $Q$ and the noisy RSS measurement, is formulated under the log-normal model. Specifically, with the unknown parameters $d$ and $\lambda$, the associated FIM, denoted $\text{FIM}(d, \lambda)$, is formulated as
\begin{equation}\label{eqn:fim}
\begin{split}
&\text{FIM}(d,\lambda)=
                   \left(
                         \begin{array}{cc}
                            \lambda f'(d)^2(\frac{1}{f(d)}+\frac{2}{S-f(d)})+\frac{\kappa}{d^2}  &  -f'(d) \\
                             -f'(d)  & \frac{2S-f(d)}{\lambda} \\
                         \end{array}
                  \right),
\end{split}
\end{equation}
where $\kappa=\left(\frac{10\alpha}{\sigma_{dB}\ln10}\right)^2$. The detailed derivation of (\ref{eqn:fim}) can be found in Appendix A.

As was proved in \cite{huang2014estimating}, $f(d)$ is a first order differentiable function, such that the CRLB for $d$ by using any unbiased estimator, denoted $\text{CRLB}(d)$, can be formulated as follows
\begin{equation}
\text{CRLB}(d)=\left(\frac{2\lambda S^2(f'(d))^2}{f(d)(2S-f(d))(S-f(d))}+\frac{\kappa}{d^2}\right)^{-1}.
\end{equation}

With the CRLB, the comparison will be made in the experimental analyses section to verify the effectiveness of the proposed method.

\section{EXPERIMENTAL ANALYSES}\label{sec:exp}
In this section, we aim to investigate the accuracy of the proposed distance estimation method, and further analyse the influences of different factors through both simulative and practical measurements.

\subsection{Simulative Analyses}
In the numerical simulations, the root-mean-square error (RMSE), which equals to the square root of the squared biases plus variances of the errors in distance estimates, is evaluated in Matlab to measure the accuracy of the proposed method. Moreover, the CRLB and the RMSE produced by the RSS-based method and the connectivity-based method are also calculated in the simulations.

The simulative parameters in relation to the WSNs and wireless channels are described below.
\begin{enumerate}
  \item The mean value of RSS measurements at the reference distance, i.e. $\overline{P_R}(d_0)$, is $-37.47$dBm;
  \item The minimum acceptable RSS value is $-100$dBm;
  \item Each WSN is deployed in a square region, the side length of which varies with different configurations of $\alpha$ and $\sigma_{dB}$;
  \item The sensors are deployed under the random and uniform distribution of mean $\lambda$;
  \item The RMSE in each case is evaluated after simulating $10000$ distance estimates;
  \item $\mu$ is the expected number of immediate neighbors of a sensor, namely $\mu=E(M+P)=E(M+Q)$, and will take different values given various configurations of $\sigma_{dB}$ and $\alpha$.
\end{enumerate}
For better presentation, the connectivity index $\mu$ will be used in the following discussions instead of the sensor density $\lambda$.

To analyze the error characteristics of the proposed method, the influences of different factors, including the expected number of immediate neighbors of a sensor, the variance of shadowing effect and the path loss exponent, are investigated in what follows.

\begin{figure*}[!htp]
  \centering
   \subfigure[RMSE ($\mu=10$)]{
  \includegraphics[scale=0.5]{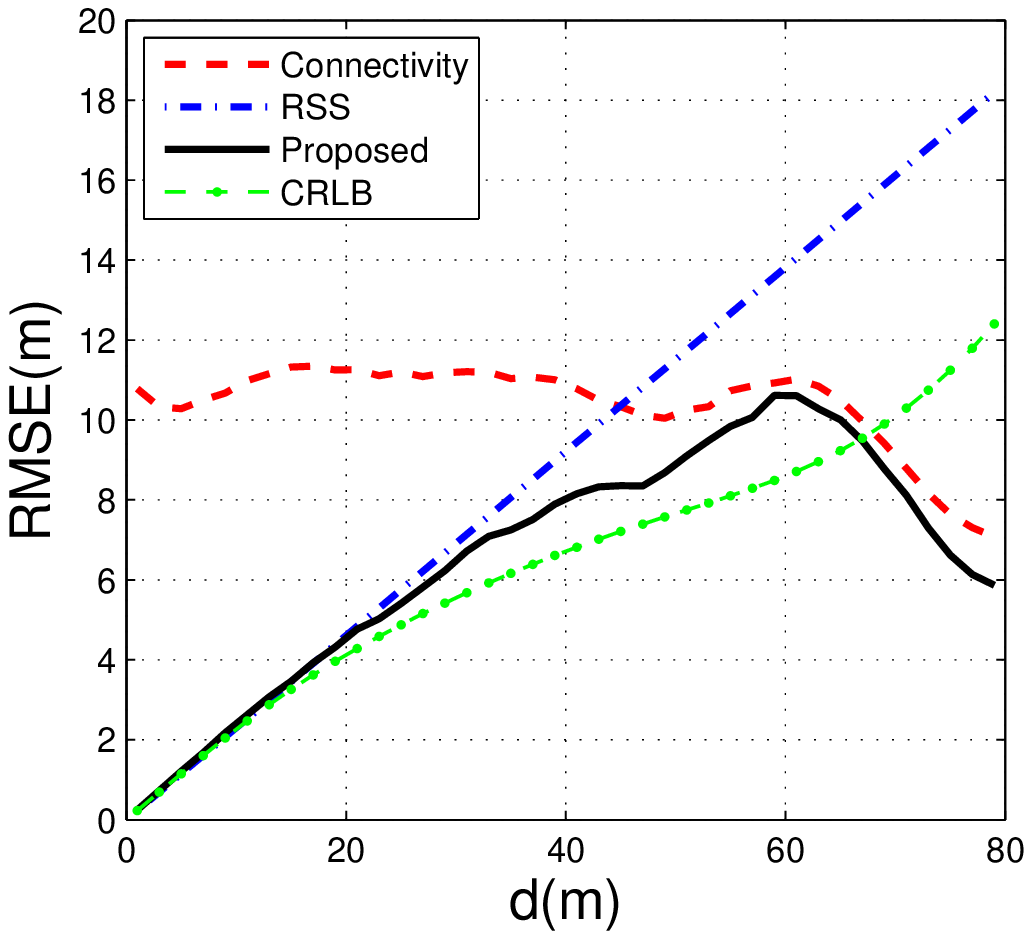}
  }
  \subfigure[RMSE ($\mu=20$)]{
  \includegraphics[scale=0.5]{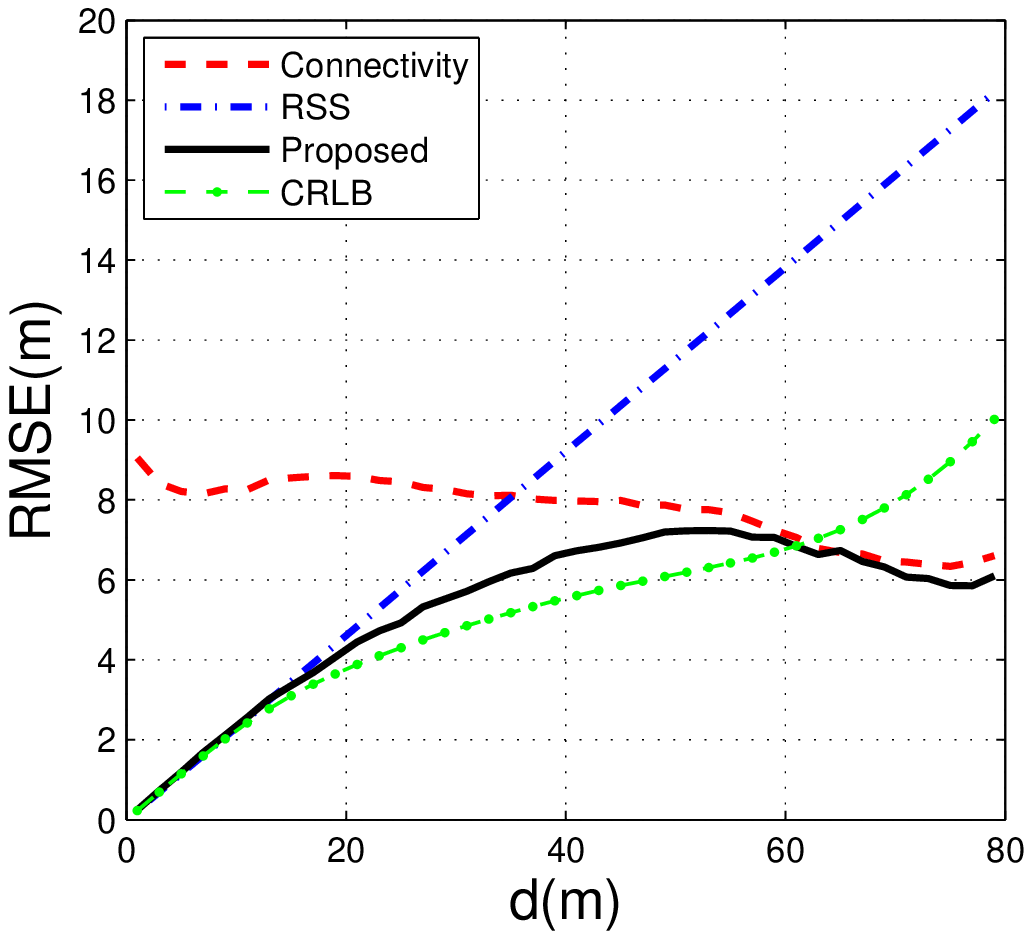}
  }
  \subfigure[RMSE ($\mu=30$)]{
  \includegraphics[scale=0.5]{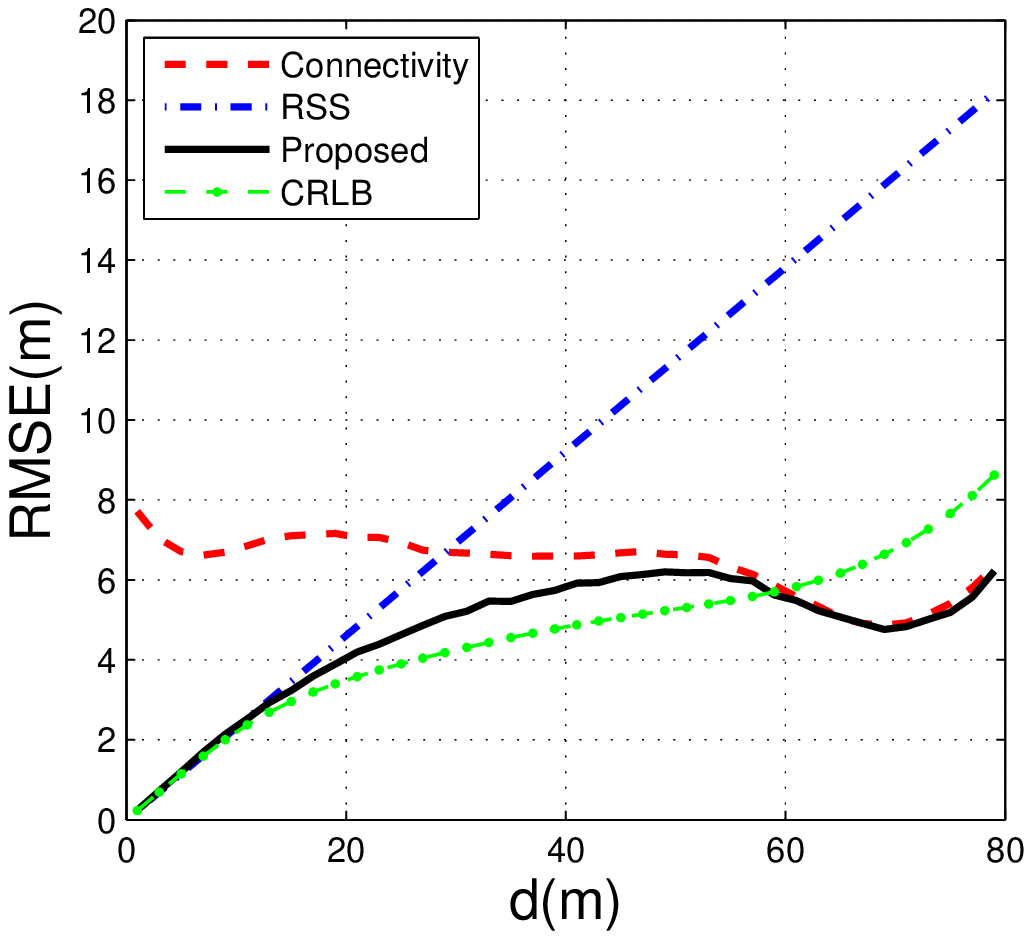}}
  \subfigure[RMSE ($\mu=40$)]{
  \includegraphics[scale=0.5]{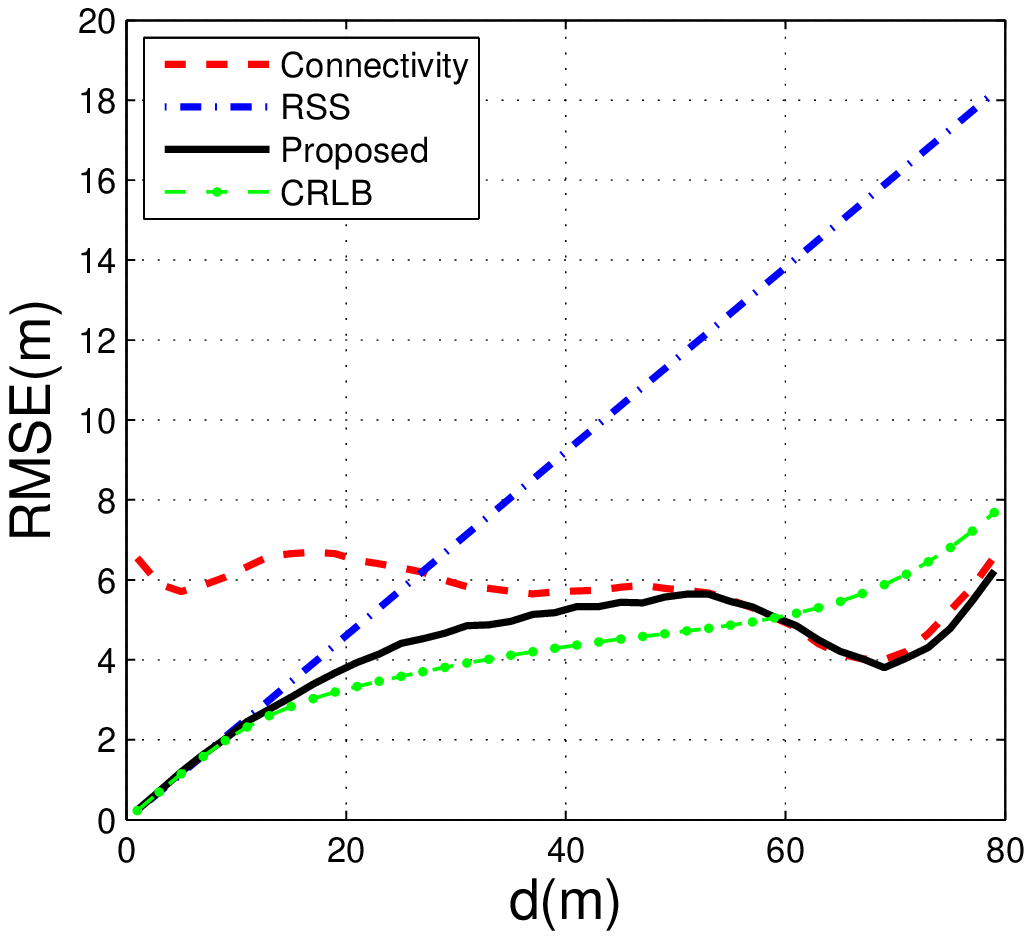}
  }
  \caption{The RMSE and CRLB given $\alpha=4$, $\sigma_{dB}=4$ and $\mu=10,20,30,40$.}
   \label{fig:difm}
\end{figure*}

Firstly, the effect of the expected number of immediate neighbors of a sensor (i.e. $\mu$) on   the RMSE is considered. Given $\sigma_{dB}=4$ and $\alpha=4$, Fig.~\ref{fig:difm} depicts the RMSE with $\mu$ varying from $10$ to $40$. As can be seen, we can observe that
 \begin{itemize}
     \item given two nearby sensors, the performance of the proposed method approaches to that of the RSS-based distance estimate, and is superior to that of the connectivity-based method; on the contrary, given two sensors far away from each other, the performance of the proposed method is close to that of the connectivity-based method and is much better than the RSS-based method; that is to say, the proposed method always outperforms the other two methods;
     \item the RMSE of the proposed method is slightly higher than the CRLB when two sensors are not far way from each other, and can even be smaller than the CRLB due to the fact that the boundary information (i.e. the upper bound on the distance estimate) is introduced.
\end{itemize}

\begin{figure*}[t]
  \centering
   \subfigure[RMSE ($\sigma_{dB}=5$)]{
  \includegraphics[scale=0.5]{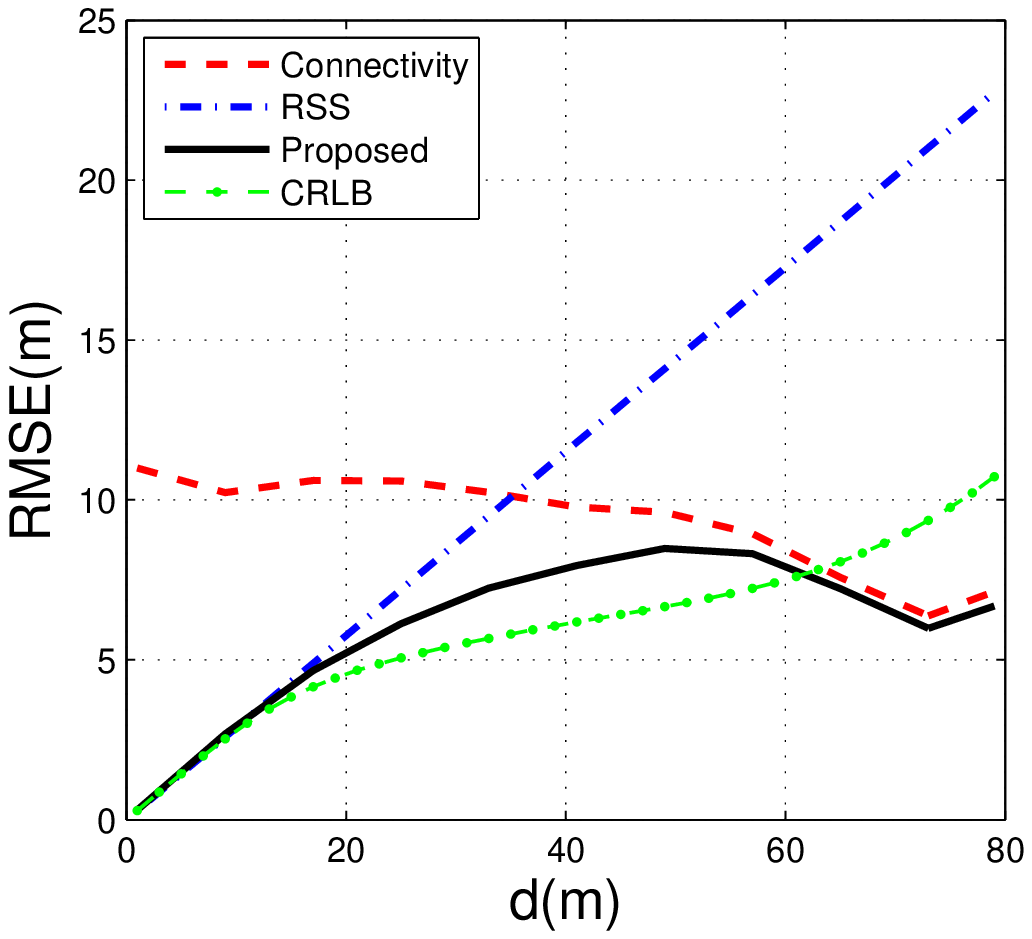}
  }
  \subfigure[RMSE ($\sigma_{dB}=6$)]{
  \includegraphics[scale=0.5]{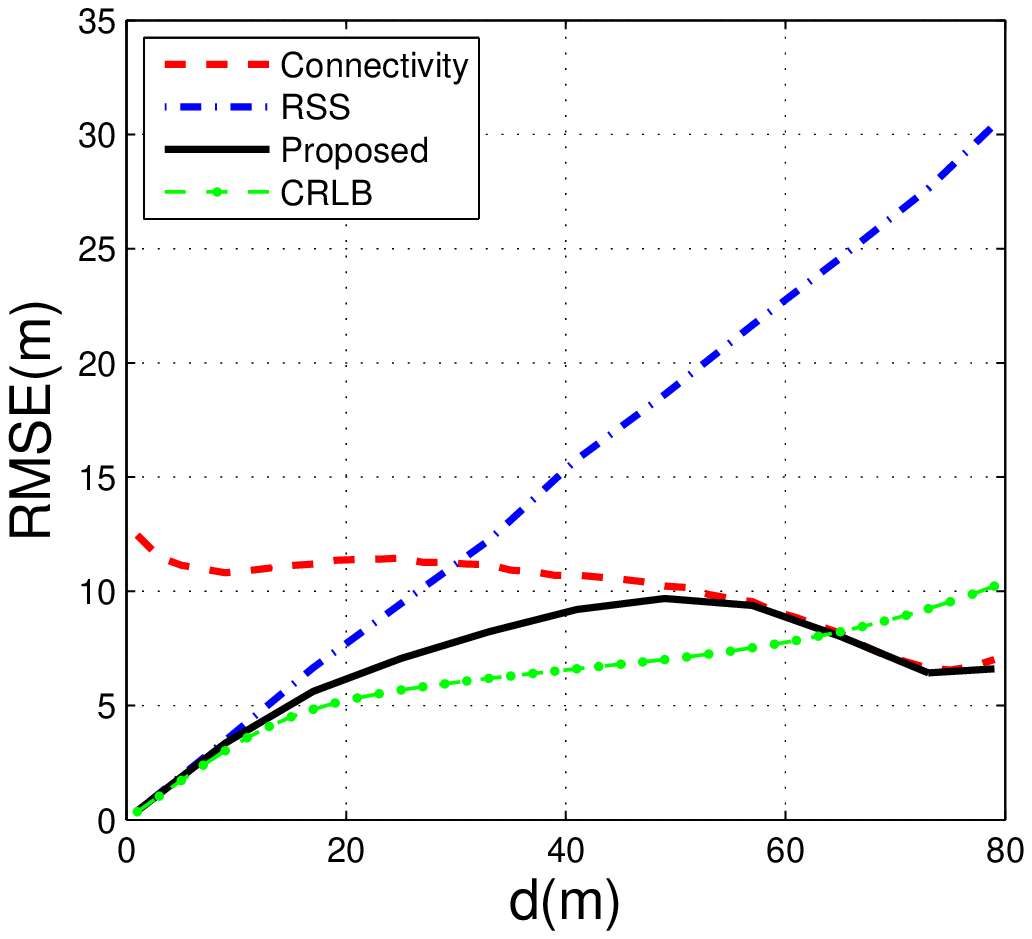}
  }
  \subfigure[RMSE ($\sigma_{dB}=7$)]{
  \includegraphics[scale=0.5]{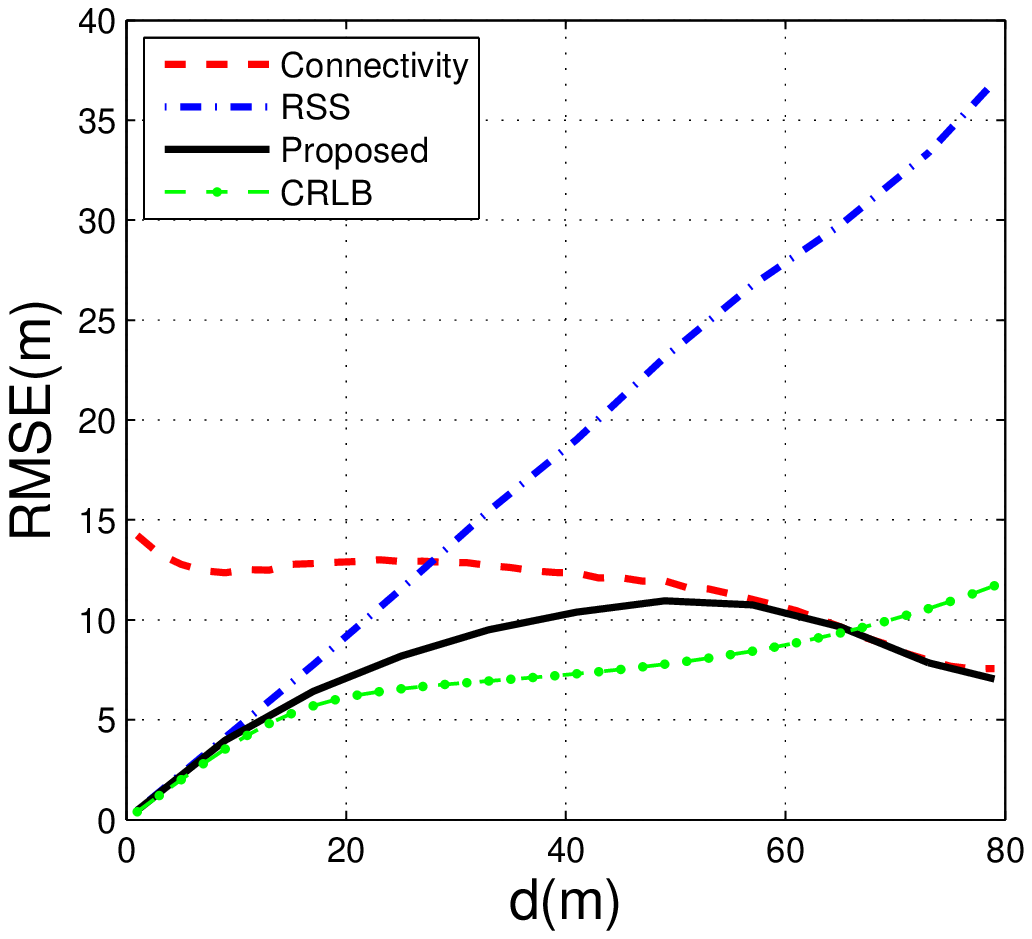}}
  \subfigure[RMSE ($\sigma_{dB}=8$)]{
  \includegraphics[scale=0.5]{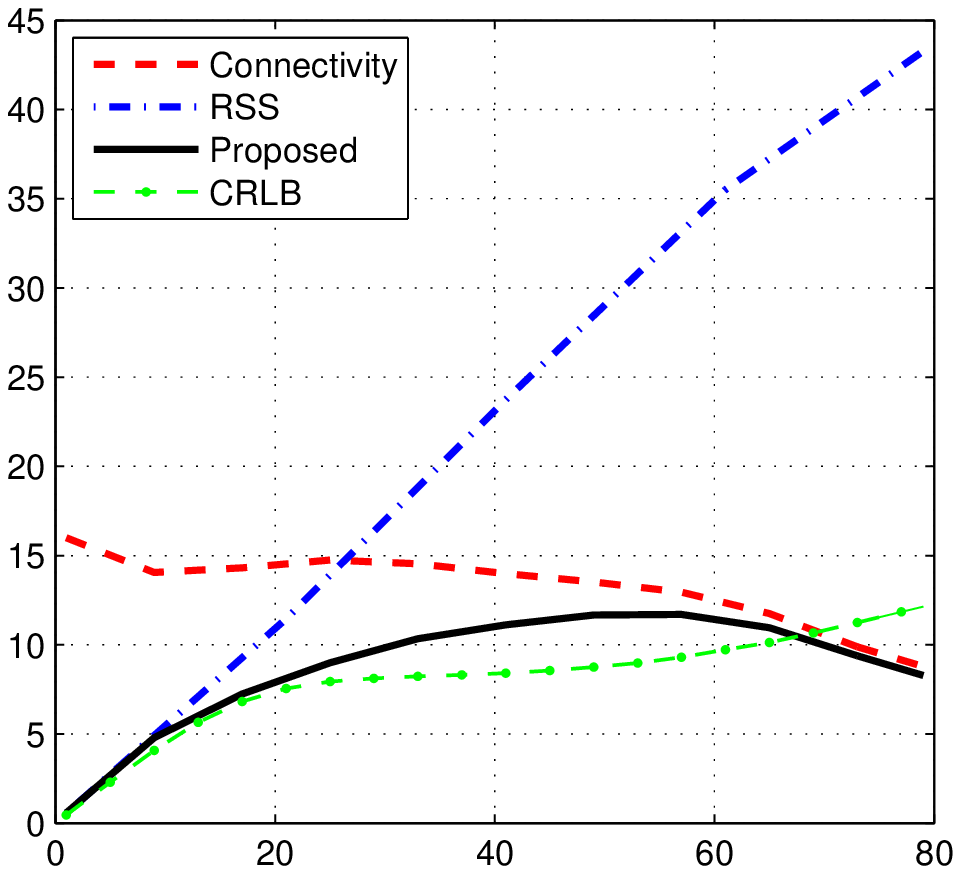}
  }
  \caption{The RMSE and CRLB given $\alpha=4$, $\mu=20$ and $\sigma_{dB}=5,6,7,8$.}
   \label{figure:difth}
\end{figure*}

Secondly, the influence of the noise level on the RMSE is investigated. As shown in Fig.~\ref{figure:difth},  the RMSE and CRLB are plotted given $\mu=20$, $\alpha=4$ and $\sigma_{dB}$ is number varying from $4$ to $8$, and it can be concluded that
\begin{itemize}
  \item with $\sigma_{dB}$ increasing, the performance of the RSS-based methods deteriorates, which is on account of the increasing noises in RSS measurements, whereas the proposed method and the connectivity-based method incur slight changes, which is also consistent with the CRLB;
  \item the overall performance of the proposed method is also better than the other two methods.
\end{itemize}

\begin{figure*}[t]
  \centering
   \subfigure[RMSE ($\alpha=3$)]{
  \includegraphics[scale=0.5]{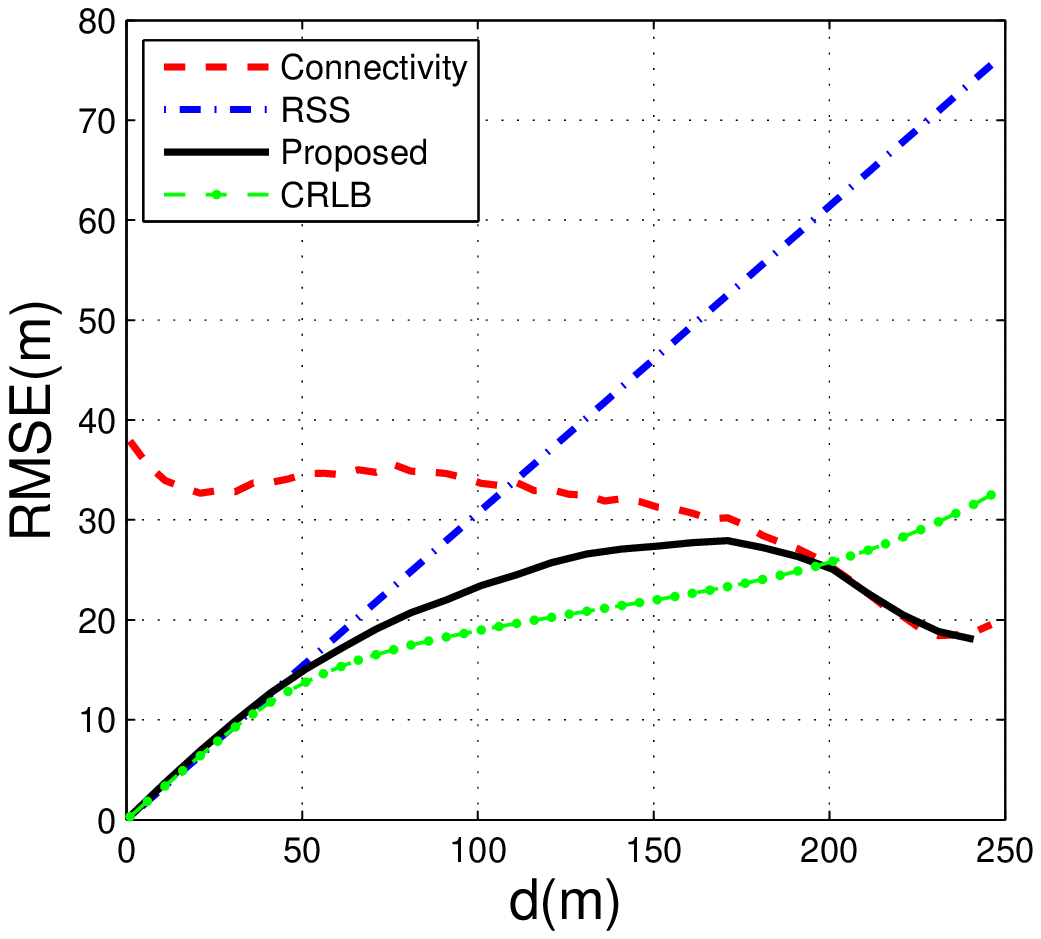}
  }
  \subfigure[RMSE ($\alpha=4$)]{
  \includegraphics[scale=0.5]{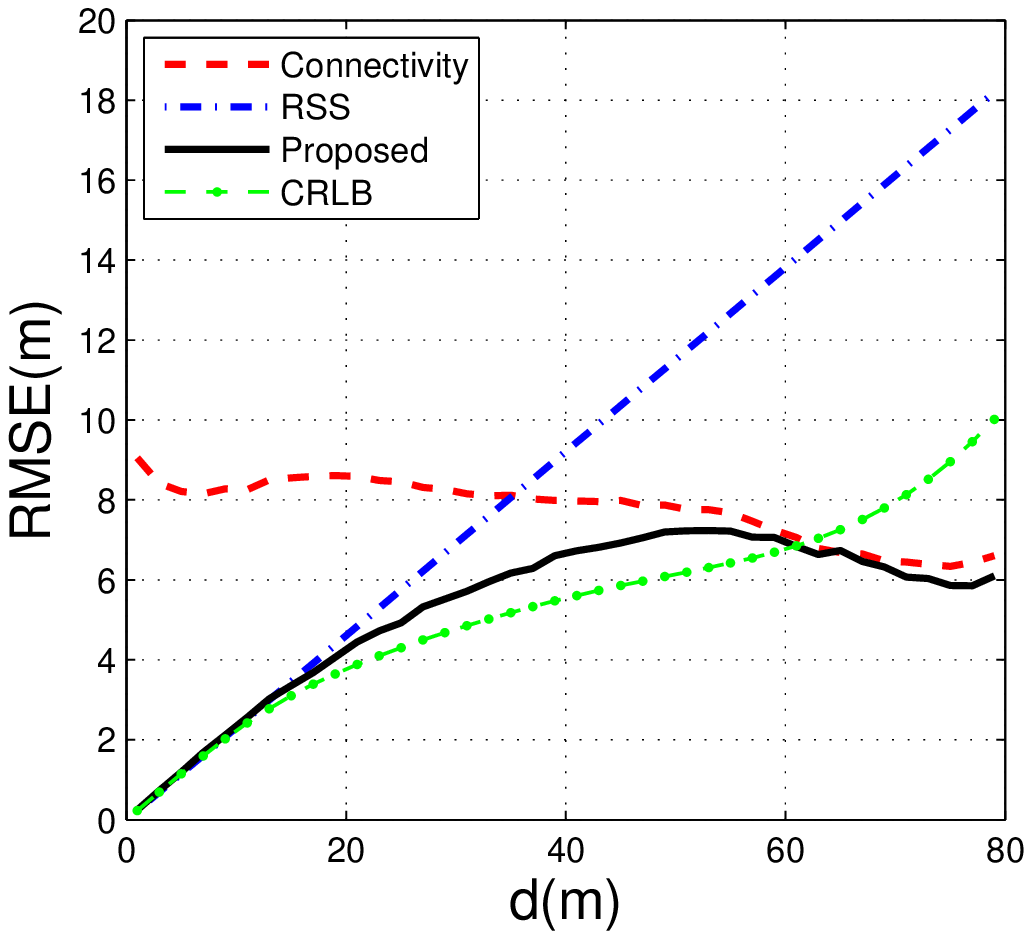}
  }
  \subfigure[RMSE ($\alpha=5$)]{
  \includegraphics[scale=0.5]{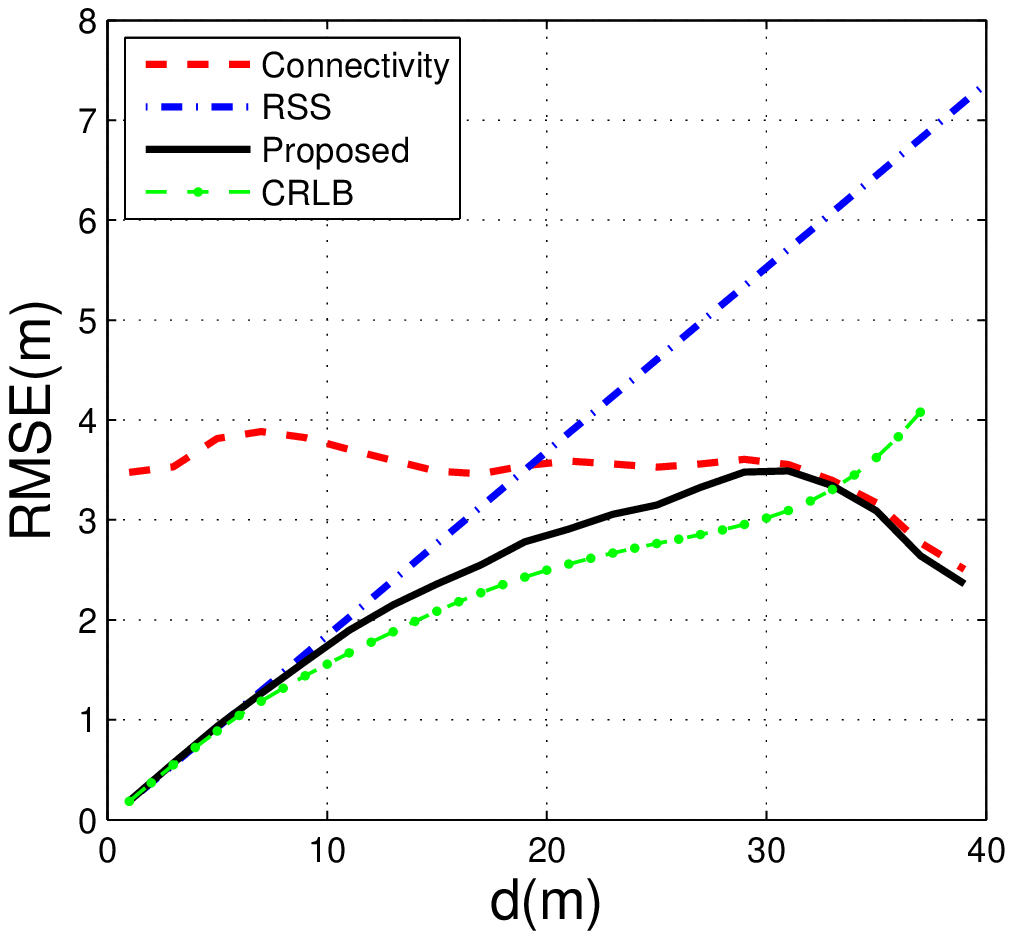}}
  \subfigure[RMSE ($\alpha=6$)]{
  \includegraphics[scale=0.5]{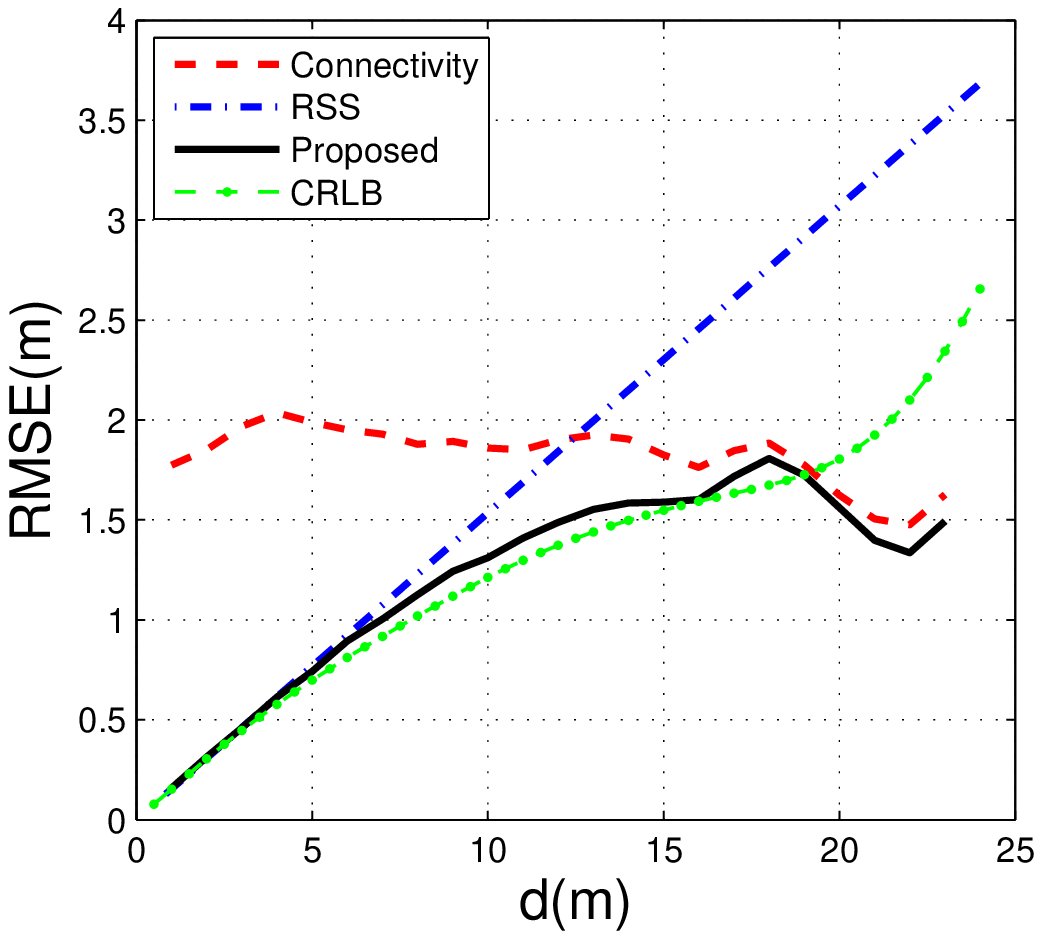}
  }
  \caption{The RMSE and CRLB given $\sigma_{dB}=4$, $\mu=20$ and $\alpha=3,4,5,6$.}
   \label{figure:difpath}
\end{figure*}

Finally, the effect of the path loss exponent is studied. As illustrated in Fig.~ \ref{figure:difpath}, the RMSE and CRLB are plotted given $\mu=20$, $\sigma_{dB}=4$ and $\alpha$ varying from $3$ to $6$. It can be seen that
\begin{itemize}
  \item with $\alpha$ increasing, both the RMSE of these three methods and the CRLB decrease, which is because the communication range decreases;
  \item the proposed method always outperforms the other two methods and approaches to the CRLB.
\end{itemize}

\subsection{Implementing the Method in Practice}

\begin{figure}[H]
  \centering
  \includegraphics[scale=0.65]{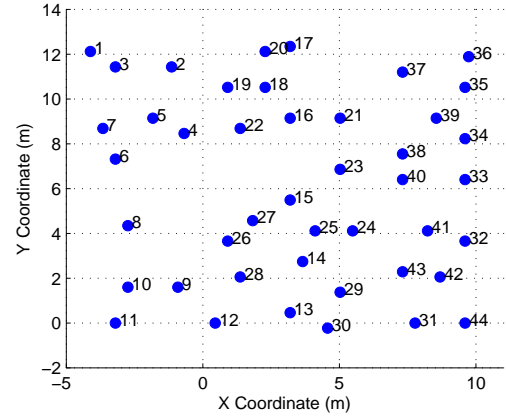}
  \caption{The layout of the sensors in \cite{Wsnlmr2011http}  }
  \label{fig:rede}
\end{figure}

To better demonstrate its superiority, the proposed method is implemented using practical RSS measurements and deployment information provided in \cite{Wsnlmr2011http}. Specifically, a WSN consisting of $44$ sensors was deployed in a real environment as shown in Fig. \ref{fig:rede}, and the RSS measurements between any two sensors were reported. On the basis of their RSS measurements, the proposed method can be run to produce practical distance estimates.

According to \cite{Wsnlmr2011http}, define $\alpha=2.3$, $\sigma_{dB}=3.92$, $\overline{P_R}(d_0)=-37.47$ dBm, and the minimum value of the RSS measurements is $-55$ dBm. To avoid boundary effects as much as possible, we consider the four sensors near the center of the deployment region, i.e. sensors $15$, $23$, $24$ and $25$. The distance estimates among these four sensors are calculated by RSS-based, connectivity-based and proposed methods, respectively, and are listed in Tab.~\ref{table:realerrors}. As can be seen, the errors of the proposed method are always less than the corresponding errors obtained by the other two methods.

\begin{table*}[t]
 \centering
  \caption{The errors by different distance estimation methods}\label{table:realerrors}
  \begin{tabular}{l|l l l l l l}
  \hline
       Sensor pair                    & (15, 23)  & (15, 24)& (15, 25)&(23, 24)& (23, 25) & (24, 25) \\ \hline
       The proposed method     & 0.603    &0.520 & 0.202 & 0.776 & 0.963 &0.313\\ \hline
       The RSS-based method  &0.974&1.168&0.442&1.419&1.654&0.745 \\ \hline
       The connectivity-based method &0.615& 0.530&0.778&0.828&1.257&0.905\\ \hline
       True distance                &2.286 & 2.656 & 1.649 & 2.781 & 2.891 & 1.371 \\ \hline
  \end{tabular}
\end{table*}

To sum up, the proposed method is able to achieve more accurate distance estimates than both the RSS-based method and the connectivity-based method under various simulative and practical environments, which confirms the effectiveness of the proposed method.

\section{CONCLUSIONS}\label{sec:con}
In this paper, we fused the RSS measurements and local connectivity between two neighboring nodes and implemented the low-cost and accurate distance estimation method. The advantages of the proposed method lie in the following aspects. Firstly, the practical log-normal model was applied to deduce the error characteristics of the RSS-based method and the connectivity-based method, which enables us to fuse two sources of information based the MLE. Secondly, both simulations and experiments were conducted, and it was shown that the proposed method outperforms its counterparts and approaches to the CRLB in most cases.

Regarding future works, we would like to apply the proposed method with the existing low-cost localization algorithms (e.g. DV-Hop) so as to improve the localization performance of WSNs without using extra devices.

\appendices
\section{The Derivation of the CRLB}

The probability density functions of $M$, $P$, $Q$ and the RSS can be formulated as follows
\begin{equation*}
\begin{split}
    &p_M(x_1)=\frac{(\lambda f(d))^{x_1}}{x_1!} \exp(-\lambda f(d)),   \\
    &p_P(x_2)=\frac{(\lambda (S-f(d)))^{x_2}}{x_2!}\exp(-\lambda(S-f(d))),  \\
    &p_Q(x_3)=\frac{(\lambda (S-f(d)))^{x_3}}{x_3!}\exp(-\lambda(S-f(d))),  \\
    &p_{P_R}(x_4)=\frac{1}{\sqrt{2\pi}\sigma_{dB}}\exp\left(-\frac{(x_4-\overline{P_R}(d_0)+10\alpha\log_{10}d)^2}{2\sigma_{dB}^2}\right),\\
\end{split}
\end{equation*}
where $P_R$ is a random variable representing RSS in dBm, and is normally distributed with the mean $\overline{P_R}(d_0)-10\alpha\log_{10}d$ and variance $\sigma_{dB}^2$.

According to the probability density functions, it is straightforward to formulate the likelihood function as
\begin{equation*}
\begin{split}
    &L(d,\lambda)=p_M(x_1)\times p_P(x_2) \times p_Q(x_3) \times p_{P_R}(x_4)\\
    \quad& =\lambda^{x_1+x_2+x_3} \frac{1}{\sqrt{2 \pi}\sigma_{dB}}\frac{(f(d))^{x_1}(S-f(d))^{x_2+x_3}}{x_1! x_2! x_3!}\\
    & \times \exp\left(-\lambda (2S-f(d))-\frac{(x_4-\overline{P_R}(d_0)+10\log_{10}d)^2}{2\sigma_{dB}^2}\right).
\end{split}
\end{equation*}
And the natural logarithm of the likelihood function can be expressed as follows
\begin{equation*}
\begin{split}
&\ln L(d,\lambda)=(x_1+x_2+x_3)\ln \lambda+\ln\left(\frac{(f(d))^{x_1}(S-f(d))^{x_2+x_3}}{x_1! x_2! x_3!\sqrt{2 \pi}\sigma_{dB}}\right)\\
    & \qquad \quad \quad -\lambda (2S-f(d))-\frac{(x_4-\overline{P_R}(d_0)+10\alpha\log_{10}d)^2}{2\sigma_{dB}^2}.\\
\end{split}
\end{equation*}

Therefore, the FIM is defined as
\begin{equation*}
\begin{split}
\text{FIM}(d,\lambda)=
-\left(
   \begin{array}{cc}
   E\left(\frac{\partial^2\ln L}{\partial d^2}\right)& \qquad E\left(\frac{\partial^2\ln(L)}{\partial \lambda\partial d}\right) \\
   E\left(\frac{\partial^2\ln(L)}{\partial \lambda\partial d}\right)& \qquad E\left(\frac{\partial^2\ln(L)}{\partial\lambda^2}\right) \\
   \end{array}
   \right).
\end{split}
\end{equation*}

Since $f(d)$ approximates a linear function, the partial derivative can be formulated as

\begin{equation*}
\begin{split}
\frac{\partial\ln L}{\partial d}=&\frac{x_1f'(d)}{f(d)}-\frac{(x_2+x_3)f'(d)}{S-f(d)}+\lambda f'(d)\\
&-\frac{x_4-\overline{P_R}(d_0)+10\alpha\log_{10}d)}{\sigma_{dB}^2}\frac{10\alpha}{d \ln10},\\
\frac{\partial\ln L}{\partial \lambda}=&\frac{x_1+x_2+x_3}{\lambda}-(2S-f(d)),
\end{split}
\end{equation*}

\begin{equation*}
\begin{split}
\frac{\partial^2\ln L}{\partial d^2}=
&-\frac{x_1(f'(d))^2}{f^2(d)}-\frac{(x_2+x_3)(f'(d))^2}{(S-f(d))^2}-\frac{\kappa}{d^2}\\
&+\frac{(x_4-\overline{P_R}(d_0)+10\alpha\log_{10}d)}{\sigma_{dB}^2}\frac{10\alpha}{d^2\ln10},\\
\frac{\partial^2\ln L}{\partial \lambda^2}=&-\frac{x_1+x_2+x_3}{\lambda^2},\\
\frac{\partial^2\ln L}{\partial \lambda \partial d}=&f'(d).
\end{split}
\end{equation*}

Because $M$, $P$, $Q$ and $P_R$ are independent random variables with
means $\lambda f(d)$, $\lambda(S-f(d))$, $\lambda(S-f(d))$ and $\overline{P_R}(d_0)-10\alpha\log_{10}d$ , respectively, we can have
\begin{equation*}
\begin{split}
    &E\left(\frac{\partial^2\ln L}{\partial d^2}\right)=-\lambda f'(d)^2\left(\frac{1}{f(d)}+\frac{2}{S-f(d)}\right)-\frac{\kappa}{d^2},\\
    &E\left(\frac{\partial^2\ln L}{\partial\lambda^2}\right)=-\frac{2S-f(d)}{\lambda},\\
    &E\left(\frac{\partial^2\ln L}{\partial \lambda \partial d}\right)=f'(d),
\end{split}
\end{equation*}
where $\kappa=\left(\frac{10\alpha}{\sigma_{dB}\ln10}\right)^2$.

Then, the FIM can be expressed as follows
\begin{equation*}
\begin{split}
&\text{FIM}(d,\lambda)=
                   \left(
                         \begin{array}{cc}
                           \lambda f'(d)^2\left(\frac{1}{f(d)}+\frac{2}{S-f(d)}\right)+\frac{\kappa}{d^2}  &  -f'(d) \\
                             -f'(d)  & \frac{2S-f(d)}{\lambda} \\
                         \end{array}
                  \right),
\end{split}
\end{equation*}
and thus, we can have
\begin{equation*}
\text{CRLB}(d)=\left(\frac{2\lambda S^2(f'(d))^2}{f(d)(2S-f(d))(S-f(d))}+\frac{\kappa}{d^2}\right)^{-1}.
\end{equation*}

\end{document}